\documentclass[aip, rsi, reprint]{revtex4-1}
\usepackage{graphics}
\usepackage{graphicx}
\usepackage{bm}
\usepackage[normalem]{ulem}

\begin{document}

\title{Partial-Transfer Absorption Imaging: A versatile technique for optimal imaging of ultracold gases}

\affiliation{Joint Quantum Institute, National Institute of Standards and Technology and University of Maryland, Gaithersburg, Maryland, 20899, USA}

\author{Anand Ramanathan}
\altaffiliation[Current address: ]{NASA/GSFC, 8800 Greenbelt Road, Mail Code 694,  Greenbelt, MD - 20771}

\author{S\'{e}rgio R. Muniz} 
\altaffiliation[Current address: ]{Instituto de F\'isica de S\~ao Carlos, Universidade
de S\~ao Paulo, S\~ao Carlos, SP 13560-970, Brazil}

\author{Kevin C. Wright}

\author{Russell P. Anderson}
\altaffiliation[Current address: ]{School of Physics, Monash University, Melbourne, Australia}

\author{William D. Phillips}

\author{Kristian Helmerson} 
\altaffiliation[Current address: ]{School of Physics, Monash University, Melbourne, Australia}

\author{Gretchen K. Campbell}
\altaffiliation[Contact email: ]{gretchen.campbell@nist.gov}


\begin{abstract}
Partial-transfer absorption imaging is a tool that enables optimal imaging of atomic clouds for a wide range of optical depths. In contrast to standard absorption imaging, the technique can be minimally-destructive and can be used to obtain multiple successive images of the same sample. The technique involves transferring a small fraction of the sample from an initial internal atomic state to an auxiliary state and subsequently imaging that fraction absorptively on a cycling transition. The atoms remaining in the initial state are essentially unaffected. We demonstrate the technique, discuss its applicability, and compare its performance as a minimally-destructive technique to that of phase-contrast imaging.  
\end{abstract}

\maketitle

  Since the realization of Bose-Einstein condensation (BEC)~\cite{BEC1, BEC2}, and later Fermi degeneracy~\cite{FermiDeg1, FermiDeg2, FermiDeg3}, in dilute atomic gases, the field of ultracold gases has expanded rapidly.  While the properties of ultracold gases are still undergoing active research, such gases are also used as tools to study a wide variety of phenomena, including those traditionally explored in condensed matter physics~\cite{RMP_UltracoldGases, ColdMoleculesBook, EsslingerFH}. Even though experiments are becoming increasingly complex and often require ever-more-sensitive optical imaging, the standard approaches to imaging remain essentially unchanged. Standard imaging techniques work \textit{optimally} for a limited range of cloud densities. Here we describe and analyze a technique, partial-transfer absorption imaging (PTAI), which was recently demonstrated~\cite{HallVortexDynamics, CriticalVelocity}, and present data illustrating the utility of this technique. PTAI enables the optimal imaging of clouds of almost any optical depth (OD). The flexibility of PTAI, combined with ease of use, makes it an important tool for imaging ultracold gases. 

In choosing an imaging technique for an ultracold atomic cloud, one traditionally has three options: absorption, fluorescence or phase-contrast imaging. Resonant imaging techniques, such as absorption and fluorescence imaging, exploit the strong interaction of ultracold gases with laser light. In these techniques, the scattering of photons by the atoms leads to heating and destruction of the sample, typically allowing only one image to be taken per sample. In addition, resonant absorption imaging cannot be used for optically thick (high column density) clouds, due to extinction of the imaging probe beam, which leads to loss of spatial and number information. Modifications such as imaging off-resonance, or imaging after expanding the cloud in time-of-flight, can overcome this limitation, but make reconstructing the original cloud profile more difficult. Fluorescence imaging works similarly to absorption imaging, but detects the scattered photons from the cloud instead of the transmitted probe beam. However, the solid angles for collecting fluorescent light are typically restricted, making it less effective, except in the case of very low optical depth~\cite{NDI1} (column density).

Phase-contrast imaging (PCI)~\cite{HuletPhaseContrast, KetterlePhaseContrast} and other dispersive techniques~\cite{HeterodyneImaging, DiffractionContrast, KetterleDarkGround} are typically used to image clouds of high (on-resonant) ODs. Such techniques use off-resonant probe light to avoid extinction of the beam and detect the phase shift due to the atom-light interaction in the transmitted probe beam. Since the scattering cross-section is reduced by imaging off-resonance, the perturbation to the sample is typically small (\textit{i.e.} minimally-destructive), allowing the cloud to be imaged multiple times. PCI is typically not used for imaging clouds of low OD as it gives a weak signal.

Using PTAI, one retains the advantages of absorption imaging for optically thin clouds while also being able to image optically thick clouds, without losing spatial and atom number information. In addition, as with PCI, PTAI can be minimally-destructive and allows for multiple images of the same sample. In this technique, a fraction of the atoms are transferred to an auxiliary internal state that has a cycling transition, and then resonantly imaged. The atoms remaining in the original state are virtually transparent to the probe light. The recoil momentum imparted by many scattered photons from the imaging light cause the transferred cloud to be ejected from the trap.  The remaining atoms, being transparent, are essentially unperturbed, stay trapped, and can subsequently be re-imaged.

PTAI is advantageous over PCI in that the degree of perturbation of the sample can be more easily varied, even for successive images of the same sample. This is useful in situations where the OD of the atomic cloud is changing significantly as a function of time or of the number of images. An example is measuring loss processes, such as when characterizing Feshbach resonances~\cite{RMPFeshbach} or Efimov states~\cite{EfimovCs, EfimovK}. PTAI could also be particularly useful for studying \textit{in situ} dynamics of quasi two-dimensional condensates where the cloud OD in the tightly confined direction may not be high enough to obtain a good phase-contrast image \cite{BKTNIST, Dalibard2DCritical, LANL_AOD}.

Any measurement on an ultracold sample perturbs it. In a minimally-destructive imaging technique, there is a trade-off between the signal-to-noise ratio (S/N) and the perturbation. For example, in PCI, decreasing the probe light detuning improves the S/N but also causes more absorption and therefore greater perturbation to the system~\cite{NDI1}. In PTAI, one transfers a certain number of atoms to the auxiliary state according to the desired S/N. One can then scatter as  many photons from the transferred atoms as one wants without causing any additional perturbation. As an aside, we note that standard resonant absorption imaging might also be made minimally-destructive~\cite{NDI1} by lowering the amount of incident light so that less than one photon per atom is scattered on average (see section \ref{Comparison_abs}).

Recently Freilich \textit{et al.}\cite{HallVortexDynamics} used PTAI to transfer atoms to an untrapped state from a magnetically trapped BEC. Since the focus of that work was to study vortex dynamics that had spatial features too small to see \textit{in situ}, the atomic distributions were imaged after expansion in time-of-flight.   In this article, we focus on using PTAI to obtain an accurate minimally-destructive image of the \textit{in situ} density profile. In addition, we study its scope and usefulness, making quantitative comparisons with PCI. There have been several other approaches used to address the limitations of standard imaging techniques~\cite{PartialRepump, StrongProbe, HallBinaryBEC}. However, such schemes have typically been destructive and could not be used to take multiple images.

The structure of the paper is as follows: In section \ref{sec:Imaging}, we discuss the technique and its applicability. In section \ref{sec:CaseExamples}, we present a demonstration of the technique. In section \ref{sec:Noise}, we discuss noise and uncertainty in the measurement and compare the performance of the technique to PCI as a minimally-destructive imaging technique. 

\section{Imaging with PTAI} \label{sec:Imaging}

In this section, we present the requirements for implementing PTAI, discuss its applicability to alkali atoms, and describe our implementation of the technique using optically trapped sodium Bose-Einstein condensates.

\subsection{Requirements for implementing PTAI}

\begin{figure}
\includegraphics[width=\columnwidth]{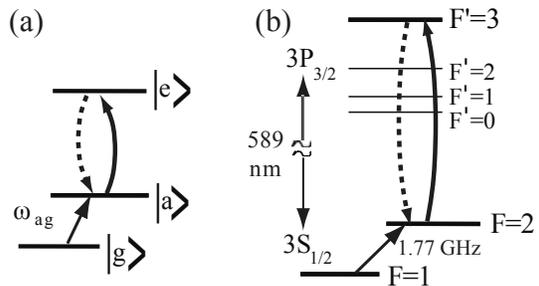}
  \caption{(a) General PTAI scheme: PTAI is implemented by transferring a small fraction of the cloud from $|g \rangle$ to $|a \rangle$ and then imaging on the cycling transition $|a \rangle$ to $|e \rangle$. (b) Sodium D$_2$ hyperfine structure : Our implementation of PTAI with $^{23}$Na uses the $|$3 $^2$S$_{1/2} $F=1$\rangle$ ($|g \rangle$) and $|$3 $^2$S$_{1/2}$F=2$\rangle$($|a \rangle$), and the $|$3 $^2$P$_{3/2} $F'=3$\rangle$($|e \rangle$) states. }
  \label{Hyperfine}
\end{figure}

PTAI can be used for any atomic species with the properties shown in Fig.~\ref{Hyperfine}(a). We assume that the atoms are initially in the atomic ground state, $|g \rangle$, from which a fraction of atoms can be transferred to the auxiliary state $|a \rangle$. The state $|a \rangle$ connects via an optical cycling transition to the excited state $|e \rangle$ for imaging. The lifetime of $|a \rangle$ must exceed imaging timescales. 

In the likely case that  $|g \rangle$ couples to a state (or states) $|e' \rangle$, the energy difference between the two transitions, $\hbar (\omega_{ea} - \omega_{e'g})$, should be large enough so that the imaging light does not cause off-resonant excitation of atoms in $|g \rangle$, \textit{i.e.}, $|\omega_{ea} - \omega_{e'g}| \gg \Gamma_{e'}$, where $\Gamma_{e'}$ is the natural linewidth of the $|g \rangle \Rightarrow |e' \rangle$ transition.

PTAI is well suited (but not limited) to imaging in optical dipole traps, which in contrast to magnetic traps, produce no spatially varying Zeeman shift that could affect the uniformity of the transfer\cite{starkshift}. Such traps are often shallow enough that the scattering of several photons transfers enough energy and momentum that the atoms leave the trap in the direction of propagation of the imaging beam. Heating due to collisions between atoms leaving the cloud and the remaining atoms is typically insignificant and was not observed by Freilich \textit{et al.}\cite{HallVortexDynamics} or in the present work. 

As with any imaging technique, the shot noise (discussed in section~\ref{sec:ShotNoise}) and technical noise in the image should be low compared to the signal (absorption in this case). In addition, in order to measure atom numbers accurately, the transfer process needs to be deterministic and reproducible.

\subsection{Using PTAI with alkali atoms}

Alkali metal atoms are used extensively in atomic physics, in part because their atomic structure allows for straightforward laser cooling. Their hyperfine structure also makes them suitable for employing PTAI. The S$_{1/2}$ ground state of alkali atoms has two hyperfine levels. Atoms can be coherently transferred between these two states using microwave or optical Raman transitions. Alkali atoms have the D$_2$ optical transition to the P$_{3/2}$ state, which allows for at least one cycling transition from the S$_{1/2}$ hyperfine state. For example, in sodium (see Fig.~\ref{Hyperfine}b), one scheme would be to assign $|g \rangle$=$|$3 $^2$S$_{1/2} $F=1$\rangle$, $|a \rangle$=$|$3 $^2$S$_{1/2} $F=2$\rangle$, and $|e \rangle$=$|$3 $^2$P$_{3/2} $F'=3$\rangle$, thereby using the cycling transition from the upper S$_{1/2}$ hyperfine state. The alternative scheme is $|g \rangle$=$|$3 $^2$S$_{1/2} $F=2$\rangle$, $|a \rangle$=$|$3 $^2$S$_{1/2} $F=1$\rangle$, and $|e \rangle$=$|$3 $^2$P$_{3/2} $F'=0$\rangle$. If there is some off-resonant excitation from $|a \rangle$ to some state $|e'' \rangle$ that decays to some state other than $|a \rangle$, or if $|a \rangle \Rightarrow |e \rangle$ is not a perfect cycling transition, and $|e \rangle$ has a decay branch to states other than $|a \rangle$, the transferred fraction will eventually be optically pumped out of the cycling transition to another state\cite{repumper}. Nevertheless, one can still use PTAI if a sufficient number of photons can be scattered before optical pumping (discussed in section~\ref{sec:UncertaintyOD}).

For either choice of $|g \rangle$-$|a \rangle$-$|e \rangle$ states in the D$_2$ transition, the other states in the P$_{3/2}$ hyperfine manifold are potential $|e' \rangle$ and $|e'' \rangle$ excited states that may affect the PTAI process via off-resonant excitation. The large energy splitting of the S$_{1/2}$ hyperfine states typically ensures that $|\omega_{ea} - \omega_{e'g}| \gg \Gamma_{e'}$ for all $|e' \rangle$ states, and there is negligible $|g \rangle \Rightarrow |e' \rangle$ off-resonant excitation. Although there is typically some $|a \rangle \Rightarrow |e'' \rangle$ off-resonant excitation, for many atomic species, the rate of optical pumping to $|g \rangle$ is low enough to use PTAI. Notable exceptions are lithium and some isotopes of potassium, which have relatively close-spaced excited levels ($\omega_{ee''} \sim \Gamma_{e''}= \Gamma_{e}$). 

In section \ref{sec:CaseExamples}, we illustrate PTAI experimentally using $^{23}$Na (Fig.~\ref{Hyperfine}b). We prepare the atoms in the $|$3 $^2$S$_{1/2} $F=1$\rangle$ ($|g \rangle$) state, and then transferred a small fraction to the $|$3 $^2$S$_{1/2} $F=2$\rangle$ ($|a \rangle$) state using a single-photon microwave process. We image the transfer fraction on the $|$3 $^2$S$_{1/2} $F=2$\rangle$ to $|$3 $^2$P$_{3/2} $F'=3$\rangle$($|e \rangle$) cycling transition~\cite{CyclingTransition}. For all $|e' \rangle$ in the $|$3 $^2$P$_{3/2}$ manifold, $|\omega_{ea} - \omega_{e'g}|/(2 \pi) >  1.6$ GHz $\gg \Gamma_{e'}/(2 \pi) =  9.8$ MHz. Optical pumping due to off-resonant excitation from the F=2 state to the F'=1,2 states ($|e'' \rangle$) followed by decay to F=1 state is low enough to allow the technique to be used.

\section{Examples of Practical Uses for PTAI} \label{sec:CaseExamples}

To illustrate the scope of the technique, we present examples of PTAI in two specific situations that have been chosen to highlight different advantages of the technique. 

\begin{figure}
  \centering
\includegraphics[width=\columnwidth]{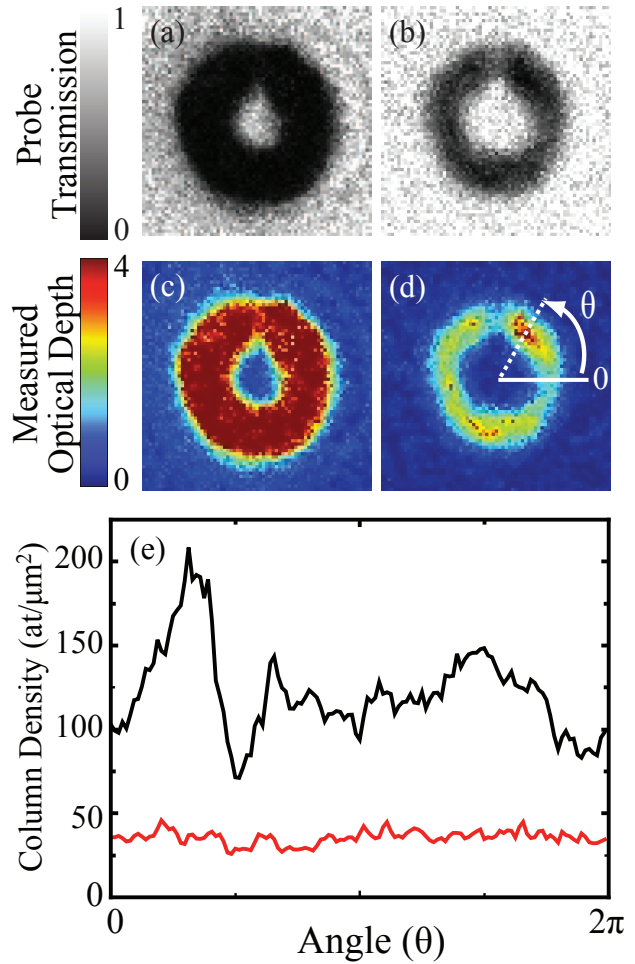}
  \caption{(Color online) Imaging an optically thick cloud: (a) Standard absorption image of an optically thick cloud (b) The corresponding PTAI image of an almost identical cloud using a fixed $\approx$15\% transfer fraction for the image. From this image we determine the initial maximum OD to be $\approx$20. (c),(d) Corresponding measured OD profiles. (e) Azimuthal column density profiles inferred from the PTAI (black line) and absorption (red line) images (angles shown in (d)). The PTAI image has been rescaled based on the known transfer fraction. Due to the severe attenuation of the probe (seen in (a), (c)), the absorption image fails to show the true OD of the cloud (see (e)) and consequently, spatial features such as the azimuthal density variation are suppressed and are disproportionately affected by shot noise (discussed in section \ref{sec:Noise}). In contrast, the PTAI image shows clear spatial features, particularly the density variations due to azimuthal inhomogeneities of the toroidal potential. All images are 85 $\mu$m $\times$ 85 $\mu$m. There are small ($<15$\%) additional corrections for saturation of the transition (both images) and optical pumping (PTAI image only). These corrections have not been made, but would not affect the overall shape of plots shown in (e).}  
  \label{HighODImaging}
\end{figure}

	\subsection{\textit{In situ} imaging of an optically thick BEC} 
\label{sec:Case1}
PTAI uses the same optical setup and probe detuning as standard absorption imaging, allowing one to easily switch between the two. In contrast, for PCI, one needs to detune the probe far from resonance and insert an optical element such as a phase-spot~\cite{KetterlePhaseContrast} or a linear polarizer~\cite{HuletPhaseContrast} to produce an interference that allows detection of the phase shift. The ease of switching between standard absorption and PTAI is advantageous in situations where for example, one may need to image a high OD cloud \textit{in situ} and a low OD cloud in time-of-flight~\cite{PCurrents, AndersonSpontaneousVortices, CriticalVelocity}, possibly even within the same experimental run. In a recent experiment~\cite{CriticalVelocity}, we created condensates in a ring-shaped trap. The cloud had an on-resonance OD of up to 20 making it unsuitable for absorption imaging \textit{in-situ} (shown in Fig.~\ref{HighODImaging}a,c), because of extinction of the probe beam. By imaging the cloud using PTAI (Fig.~\ref{HighODImaging}b,d), where we transferred $\approx$15\% of the atoms to the imaging state, we are able to observe the density variations due to inhomogeneities in the azimuthal potential (Fig.~\ref{HighODImaging}d). Plotting the azimuthal profile (Fig.~\ref{HighODImaging}e), one can see that the PTAI image (black line) clearly shows the full 200 atoms/$\mu$m$^2$ column density and the nearly 50\% density variation, while standard absorption imaging (gray line) saturates around 40 atoms/$\mu$m$^2$ and does not see the true column density.

	\subsection{Accounting for varying initial condition}
\label{sec:Case2}

Creating an ultracold gas sample takes anywhere from the order of a second to the order of a minute (for the experiments described here, it is around 30 s). Conditions may vary from sample to sample causing fluctuations in the number of atoms and consequently other properties such as the condensate fraction. Typically, the creation of the ultracold gas sample is just the starting point of a more complicated experiment. Although it is possible to average over many experimental realizations to overcome the fluctuations in the initial sample, it is easier to make a minimally-destructive measurement of the initial sample and normalize appropriately. This kind of procedure is used to study, for example, photoassociation~\cite{NaPhotoassociation} or Feshbach resonances~\cite{FeshbachRoberts}.

In order to compensate for the initial fluctuations, the uncertainty in the measurement has to be smaller than those fluctuations. In one example of this procedure, we have used PTAI to compensate for atom number fluctuations to obtain a more precise measurement of the lifetime of a BEC in a trap. Although here we have discussed normalization of atom number, one can also use a PTAI image to compensate for other fluctuations such as changes in the shape and position of the sample~\cite{PCurrents}, for example to measure trap oscillations~\cite{KetterleCollectiveExcitations}.

\section{Measurement Uncertainty} \label{sec:Noise}

In this section, we derive the measurement uncertainty of PTAI using a model calculation and simplifying assumptions. We take the relative uncertainty of the measured OD to be the important figure of merit and compare this uncertainty for PTAI and PCI for clouds of various ODs.

\subsection{Formalism}

In order to examine the measurement uncertainty, we first set up a generalized formalism to analyze the PTAI process. The transmitted probe beam is imaged on a CCD, which consists of a two-dimensional array of pixels (photosensors). Each pixel on the CCD receives probe light transmitted through a specific area of the ultracold gas cloud. For simplicity, we assume that the optical resolution is better than our pixel size and hence can be ignored for this analysis. Similarly, we assume that the imaging depth of field is larger than the relevant cloud dimension and does not affect the measurement. The measurement uncertainty depends in part on the pixel size, and is ultimately determined by shot noise. One can use a larger pixel size (by binning over adjacent pixels or reducing the optical magnification) to lower shot noise at the expense of spatial resolution. 

\begin{figure}[!ht]
  \centering
\includegraphics[width=0.3\textwidth, angle=-90]{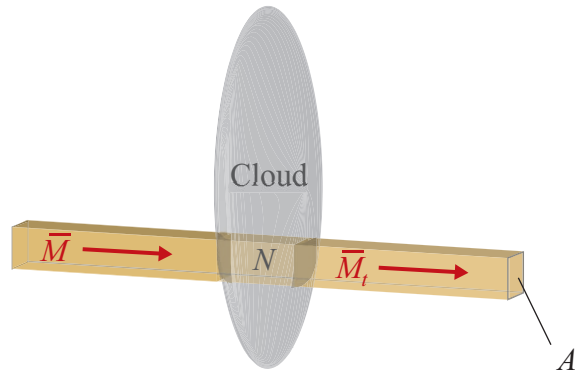}
  \caption{(Color Online) Analysis scheme: An average of $\bar{M}$ photons of the incoming probe beam within cross-section $A$ are incident on the cloud and pass through the enclosed volume, which contains $N$ atoms. The transmitted $\bar{M}_t$ photons are ultimately incident on a single detector element (pixel) of a two-dimensional array of photosensors (CCD). For simplicity, we ignore the imaging system.}  
  \label{ImagingCell}
\end{figure}

In the following analysis, we consider a part of the cloud with cross-section area $A$, containing $N$ atoms (see Fig.~\ref{ImagingCell}) in the volume enclosed by $A$ along the propagation direction. The statistical fluctuations in $N$ are ignored in this analysis. A probe pulse of duration $\tau$ passes through the cloud and is incident on the CCD. An average of $\bar{M}$ photons are incident on the imaging area $A$. $\bar{M}_t$ photons are transmitted through this area and ultimately fall on a single effective pixel, which is our detector. 

We assume that the probe intensity, $\hbar \omega \bar{M}/ (A \tau)$, where $\omega$ is the frequency of the light, is much lower than the saturation intensity of the transition. The probe time $\tau$ is short enough that atoms are assumed stationary (i.e. they move a small distance compared to $\sqrt{A}$, due to the velocity imparted to the atoms by recoil from absorption and emission of photons). We assume a single detection efficiency coefficient $\eta$ for the imaging system, which includes both the transmission efficiency of the optical beam path to the CCD and the quantum efficiency of the detector. We also assume that the detector is shot noise limited. 

For the probe pulse of frequency $\omega$, imaging on a cycling transition with resonance frequency $\omega_0$, and linewidth $\Gamma$, we define a normalized detuning $\Delta = (\omega - \omega_0)/(\Gamma/2)$. We ignore broadening from other effects, which are typically negligible compared to $\Gamma$ for ultracold gases. The (on-resonant) OD of the area of interest is given by:
\begin{equation}
	\beta = N \sigma_0 / A,
\end{equation}
where $\sigma_0$ is the resonant scattering cross-section.  We assume that the atom cloud density varies slowly over distances on the order of the wavelength of the light, and obtain\cite{KetterleReview}:
\begin{eqnarray}
\bar{M}_t &=& \bar{M} \exp\left( -\frac{\beta}{1+ \Delta^2} \right) \label{eq:transmission}, \\
\bar{M}_\mathrm{abs} &=& \bar{M}- \bar{M}_t = \bar{M} \left[ 1- \exp\left( -\frac{\beta}{1+ \Delta^2} \right) \right] \label{eq:absorption}, \\
\delta \phi &=& \beta \frac{\Delta}{2(1+ \Delta^2)}, \label{eq:phaseshift}
\end{eqnarray}
where $\bar{M}_\mathrm{abs}$ is the average number of absorbed photons and $\delta \phi$ is the phase shift imparted to the transmitted probe by the atoms. For PTAI, imaging is best done on resonance ($\Delta=0$) for which there is no phase shift ($\delta \phi$=0). 

In using the PTAI technique, we transfer a chosen fraction of atoms, $f$, corresponding to an average transferred OD of $f \beta$, and number $f N$. In this calculation, we assume that atoms are neither optically pumped out of the cycling transition nor mechanically pushed out of $A$ during the imaging process. In practice, this puts a limit on the number of incident photons $\bar{M}$, which will be discussed later in this section. With this description, we can now proceed to calculate the uncertainty of the measurement.

\subsection{Shot noise}
\label{sec:ShotNoise}
Shot noise limits the precision of PTAI in two ways, the photon shot noise of the light and the quantum projection noise of the transferred atoms. The detector photon count depends on the light transmitted through the sample, $\bar{M}_{t}$ and the detection efficiency, $\eta$. The photon shot noise $\delta \mathcal{N}_\mathrm{ph}$ of the beam is given by the square-root of the mean number of detected photons:
\begin{equation}
  \delta \mathcal{N}_\mathrm{ph} = \sqrt{\eta \bar{M}_t}= \sqrt{\eta \bar{M} e^{-f \beta}} = \sqrt{\eta \bar{M} e^{- fN \sigma_0 /A}}, \label{eq:PAIPhotonShotNoise}
\end{equation}
where we have taken $\Delta=0$.

 We view the partial transfer as placing each atom in the cloud in a coherent superposition of $|g \rangle$ and $|a \rangle$. The imaging pulse collapses the superposition into an incoherent mixture. The fluctuations from this quantum projection give a standard deviation of $\sqrt{f N}$ for a small transfer fraction $f$ of $N$ atoms. We express this fluctuation in terms of the variation $\delta \mathcal{N}_\mathrm{qp}$ in photon counts on the detector. 

\begin{eqnarray}
  \delta \mathcal{N}_\mathrm{qp} &=&  \frac{d (\eta \bar{M}_t)}{d (f N)} \sqrt{fN}  \nonumber \\
  \delta \mathcal{N}_\mathrm{qp} &=& \eta \bar{M} e^{-f \beta} \sqrt{\frac{\sigma_0 f \beta}{A}}.
\end{eqnarray}

Since the noise sources are independent, they add in quadrature,
\begin{eqnarray}
  \delta \mathcal{N} &=& \sqrt{\delta \mathcal{N}_\mathrm{ph}^2 + \delta \mathcal{N}_\mathrm{qp}^2}, \nonumber \\ 
     &= & \sqrt{\eta \bar{M} e^{-f \beta} \left[1 + \eta \bar{M} e^{-f \beta} (f \beta \sigma_0 /A) \right]} .
\end{eqnarray}

For a given $\beta$ and small $f$, PTAI gives a poor measurement of $\bar{M}_\mathrm{abs}$, that is, a low S/N =$\bar{M}_\mathrm{abs}/ \delta \mathcal{N}$, and therefore gives a poor measurement of $\beta$. While the S/N in $\bar{M}_\mathrm{abs}$ improves with larger $f$ with the best S/N at complete transfer ($f = 1$), this does not necessarily lead to an improved measurement of $\beta$. For optically thin clouds ($f \beta \lesssim$1), increasing $f$ does improve the measurement of $\beta$. However, when $f \beta>$4, changes in $\beta$ produce little change in the transmitted intensity because the probe is almost completely absorbed, a feature not accounted for by considering the S/N of $\bar{M}_\mathrm{abs}$. We therefore choose the fractional uncertainty in the measured OD $\beta$ as the metric for the quality of the image, which we will now derive by propagating $\delta \mathcal{N}$ through the image analysis.


	\subsection{Uncertainty in measured Optical Depth}
\label{sec:UncertaintyOD}
In determining the fractional uncertainty of $\beta$, an important consideration is the effect of optical pumping, which transfers atoms out of the cycling transition ($|a \rangle$--$|e \rangle$). To minimize optical pumping effects, we restrict $\bar{M}$ by putting a limit on the number of photons each atom will absorb (and emit):
\begin{eqnarray}
   \frac{\bar{M}_\mathrm{abs}}{f N} &\leq& M_\mathrm{p} , \nonumber \\
   \bar{M} &\leq& \frac{M_\mathrm{p} f N}{1- e^{-f \beta}}, \label{eq:photonlimit}
\end{eqnarray}
where $M_\mathrm{p}$ is the number of photons an atom can absorb before a chosen fraction of atoms are optically pumped. Here, we choose to limit the optically pumped fraction to 10\% for the comparisons given below. $M_\mathrm{p}$ is species specific, depending on the atomic states that contribute to optical pumping and the transition probabilities associated with those states.

Setting $\bar{M}$ to the maximum value allowed by equation (\ref{eq:photonlimit}), the total noise, $\delta \mathcal{N}$, is then
\begin{equation}
  \delta \mathcal{N} = \sqrt{\frac{M_\mathrm{p} f N \eta e^{-f \beta}}{1- e^{-f \beta}} \left( 1 + \frac{M_\mathrm{p} \eta (f \beta)^2 e^{-f \beta}}{1-e^{-f \beta}}.    \right)} \label{eq:totalnoise}
\end{equation}
The measured OD is determined by comparing the image against a reference image taken in the absence of atoms, as in traditional absorption imaging\cite{KetterleReview}. Here, we assume for simplicity, that the reference is averaged over several images and has no shot noise associated with it.

The \textit{measured} OD is, on average, $f \beta$. From this, and the chosen transfer fraction $f$, we calculate the original OD $\beta$. The measurement uncertainty, $\delta \beta/ \beta$, can be calculated using equation~\ref{eq:totalnoise} (see appendix \ref{sec:PartialTransferCalc}):
\begin{equation}
 \label{eq:PTdBr}   \frac{\delta \beta}{\beta} = \sqrt{\frac{(1-e^{-f \beta}) +M_\mathrm{p} \eta (f \beta)^2 e^{-f \beta} } {M_\mathrm{p} \eta (f \beta)^3 (A/\sigma_0) e^{-f \beta} } } ,
\end{equation}
where the only experimental variables are $f \beta$ and the imaging area $A$. 

Equation (\ref{eq:PTdBr}) illustrates how the photon shot noise and quantum projection noise affect $\delta \beta/ \beta$. In equation (\ref{eq:PTdBr}), the first term $(1-e^{-f \beta})$ in the numerator arises from the photon shot noise, while the second term arises from the quantum projection noise.  For sufficiently low optical pumping ($M_\mathrm{p} \eta \gtrsim 10$), photon shot noise plays a dominant role only in two limits: small $f \beta$ such that $M_\mathrm{p}\eta f \beta < 1$ and large $f \beta $ such that $M_\mathrm{p} \eta (f \beta)^2 e^{-f \beta} < 1$. 


The former limit corresponds to a transferred cloud that is optically thin where few photons are absorbed because of the limit on $\bar{M}$ due to optical pumping. In this limit, increasing $f$ improves the signal and therefore lowers the fractional uncertainty. The latter limit corresponds to a transferred cloud that is optically thick, where the transmitted probe is nearly extinguished, causing the measurement to be limited by photon shot noise. Here, increases in $f$ attenuate the probe further causing the fractional uncertainty $\delta \beta/ \beta$ to increase.

\begin{figure}
  \centering
\includegraphics[width=\columnwidth]{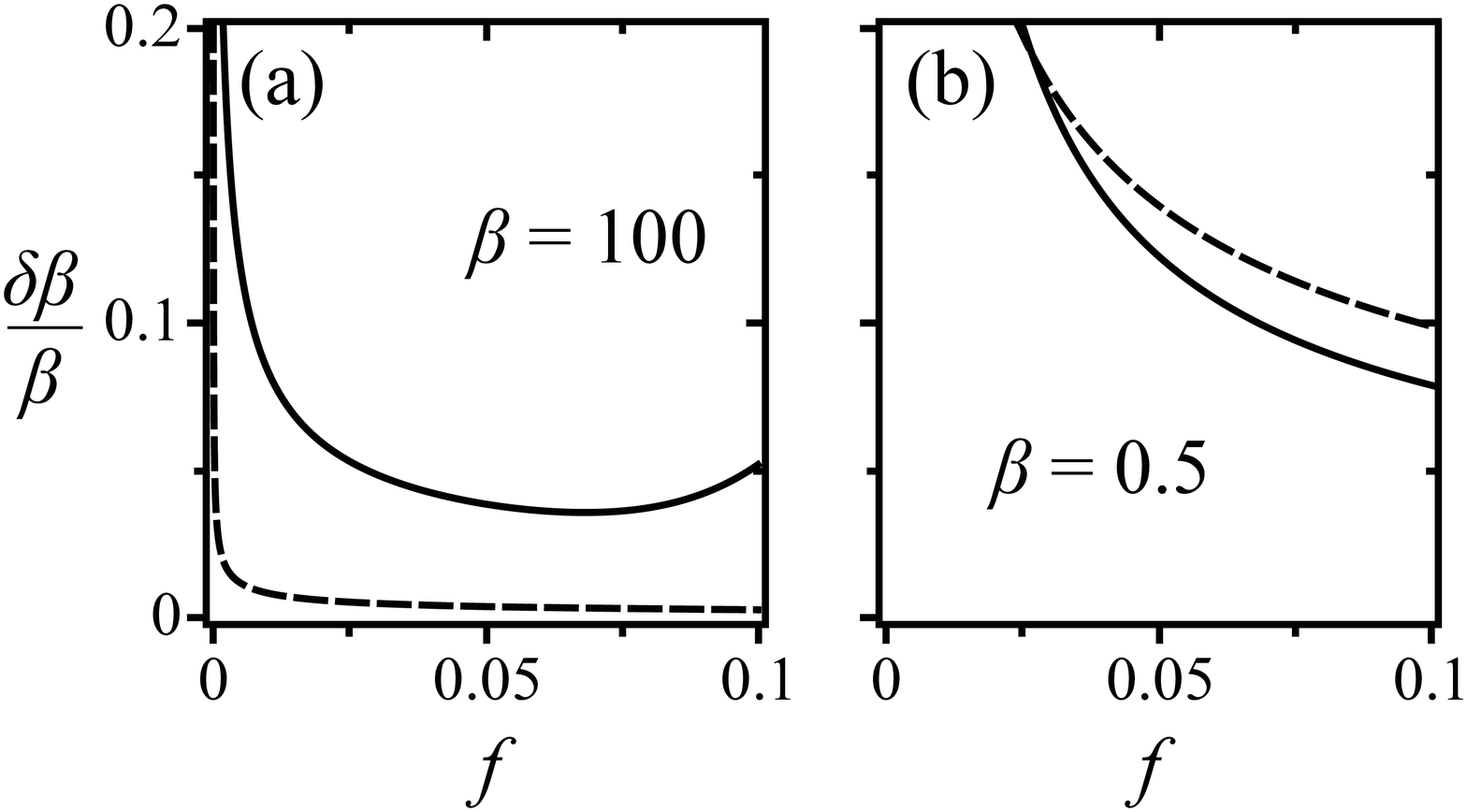}
  \caption{Measurement uncertainty vs fractional atom loss  : We calculate the uncertainty of PTAI (from equation (\ref{eq:PTdBr})) and PCI (equation (\ref{eq:PCdBrSimpl2})), and compare them assuming $f_\mathrm{r}$ to be equivalent to $f$.  (a) For a optically thick cloud, $\beta=100$, PCI (dashed) gives a lower uncertainty than the PTAI technique (solid). $A = 1.5 \times 1.5 \mu$m$^2$. (b) For an optically thin cloud, $\beta=0.5$, PTAI (solid) works better than PCI (dashed). To achieve atom numbers comparable to that of the optically thick cloud shown in (a), the imaging area in (b) is a factor of 200 larger ($A \approx 20 \times 20 \mu$m$^2$). For both techniques, $\delta \beta /\beta$ decreases with increasing $f$ for $f\ll1$, showing the trade-off between measurement uncertainty and perturbation of the sample. However, for PTAI at large $\beta$, the uncertainty reaches a minimum (as seen in (a)), before increasing with higher transfer fractions due to attenuation of the probe beam (high $f \beta$). The PCI detuning is chosen so that $\Delta^2 \gg 1$ and the phase-shift is modest ($\delta \phi < \pi/4$). In both plots, for PTAI, we use $M_\mathrm{p}=75$, the approximate value for our sodium atom experiments, and we have set $\eta=1$.} 
  \label{ODLowHigh}
\end{figure}

Between the two limits,  $\delta \beta/ \beta$ goes through a minimum as can be seen in Fig.~\ref{ODLowHigh}a (solid line). The exact position of the minimum in $\delta \beta/\beta$ depends on $M_\mathrm{p}$. For a given $M_\mathrm{p}$, equation (\ref{eq:PTdBr}) can be used to find the optimum $(f \beta)$ such that $\delta \beta/\beta$ is minimized.

PTAI works best when it is quantum projection noise limited, \textit{i.e.} in the intermediate regime between the two limits described above where photon shot noise dominates. Neglecting photon shot noise, the terms containing $M_\mathrm{p}$ dominate in equation (\ref{eq:PTdBr}), giving
\begin{equation}
 \label{eq:PTdBrSimpl} \frac{\delta \beta}{\beta} =  \sqrt{\frac{1} {f \beta (A/\sigma_0) }} =  \frac{1}{\sqrt{f N}} ,
\end{equation}
where $f N$ corresponds to the number of atoms transferred and hence lost due to the imaging process. This expression is valid only for small $f$. Equation (\ref{eq:PTdBrSimpl}) highlights the minimally-destructive nature of PTAI, where the uncertainty in measurement decreases with increasing perturbation ($f$) to the sample and as expected, is the inverse of the square root of the number of atoms lost.



\subsection{Advantages over standard Absorption Imaging}
\label{Comparison_abs}
Standard absorption imaging using low intensities of probe light can also be used as a minimally-destructive technique~\cite{NDI1}. To see the benefits of PTAI over absorption imaging (in this minimally-destructive regime), we make a comparison of the two. Following a procedure similar to that of the previous section, we obtain:
\begin{equation}
 \label{eq:AbsdBr}   \frac{\delta \beta}{\beta} = \sqrt{\frac{1-e^{-\beta}} {\eta \beta^2 f_\mathrm{r} N e^{-\beta} } },
\end{equation}
where $f_\mathrm{r} \ll 1$, the fraction of atoms that undergo photon recoil due to absorption, is a measure of the perturbation to the sample by the imaging process~\cite{N_fvsN_d}. This expression differs from equation (\ref{eq:PTdBr}) in that it is completely dependent on the photon shot noise.  

In the low $f \beta$ limit,  where PTAI is also photon shot noise limited ($M_\mathrm{p}\eta (f \beta)^2 \ll 1$), equation (\ref{eq:PTdBr}) can be simplified:
\begin{equation}
 \label{eq:PTdBrLowBeta} \left(\frac{\delta \beta}{\beta}\right)_\mathrm{PTAI} = \sqrt{\frac{1-e^{-f \beta} } {M_\mathrm{p} \eta (f \beta)^2 f N e^{-f \beta }} } 
                                                                    \approx \sqrt{\frac{1}{M_\mathrm{p} \eta f^2 \beta N}} .
\end{equation} 
This fractional uncertainty is identical to the corresponding expression in the low $\beta$ limit for (minimally-destructive) standard absorption imaging,
\begin{equation}
 \label{eq:AbsdBrLowBeta} \left(\frac{\delta \beta}{\beta}\right)_\mathrm{Abs} \approx \sqrt{\frac{1}{\eta \beta f_\mathrm{r} N}},
\end{equation} 
except for a factor of $(M_\mathrm{p} f)^{-1/2}$, which arises due to the use of the cycling transition in PTAI and allows one to scatter many photons per atom. Typically, as we observed with sodium, $M_\mathrm{p} f > 1$, which indicates that PTAI will have a lower fractional uncertainty for the same perturbation.

At higher $\beta$, $\delta \beta / \beta$ for absorption imaging grows rapidly with $\beta$ due to the $e^{-\beta}$ term in the denominator as seen in equation (\ref{eq:AbsdBr}), causing it to continue to perform worse relative to PTAI. Finally, in the limit of an optically thick cloud, $\beta>4$, this $e^{-\beta}$ term leads to such a high $\delta \beta / \beta$ that absorption imaging cannot be used (as seen in Fig.~\ref{HighODImaging}).

\subsection{Comparison with Phase-Contrast Imaging}
\label{sec:Comparisons}
We now compare PTAI with PCI as a minimally-destructive technique by comparing $\delta \beta/\beta$ for a given perturbation, \textit{i.e.} atoms lost from the sample. The corresponding expression to equation (\ref{eq:PTdBr}) for the uncertainty of the phase-contrast imaging process is (see appendix \ref{sec:PhaseContrastCalculations})
\begin{equation}
   \delta \beta = \frac{1}{\cos{\left(\frac{\beta}{2 \Delta}\right)}} \sqrt{\frac{\beta}{\eta f_\mathrm{r} N}}, \label{eq:PCdBrSimpl}
\end{equation}
where $f_\mathrm{r} \ll 1$ is the fraction of atoms that absorb a photon during the imaging process, and $\Delta$ is the normalized detuning of the off-resonant probe beam. Typically, one chooses $\Delta$ such that the phase shift of the transmitted probe is small ($\beta/\Delta \ll 1$ in equation (\ref{eq:phaseshift})). Simplifying equation (\ref{eq:PCdBrSimpl}), we find 
\begin{equation}
   \frac{\delta \beta}{\beta} = \sqrt{\frac{1}{\eta \beta f_\mathrm{r} N}} .
\label{eq:PCdBrSimpl2}
\end{equation}
As with PTAI (and minimally-destructive standard absorption imaging), the perturbation of the sample can be quantified in terms of $f_\mathrm{r}$, the fraction of atoms lost due to recoil in the imaging process~\cite{N_fvsN_d}.

Comparing equations (\ref{eq:PTdBrSimpl}) and (\ref{eq:PCdBrSimpl2}), we see that both techniques have $\delta \beta / \beta \propto 1/\sqrt{f}$. However, their $\beta$ and $\eta$ dependence differ in that $\delta \beta / \beta$ for PCI has an additional $1/ \sqrt{\eta \beta}$. Hence, for $\eta \beta > 1$, PCI gives a lower uncertainty for a given perturbation (Fig.~\ref{ODLowHigh}(a)), while at low OD ($\eta \beta<1$), as long as the uncertainty in $f$ due to technical noise is small ($<8$\% for the case considered in Fig.~\ref{ODLowHigh}(b)), PTAI gives a lower uncertainty.

\begin{figure}
  \centering
\includegraphics[width=0.95 \columnwidth]{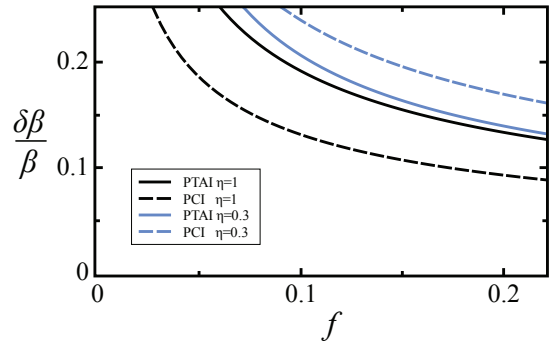}
  \caption{(Color online) Lower sensitivity of PTAI to imaging losses: For $\beta=2$, PCI (dashed) gives a lower uncertainty than the PTAI technique (solid) in the absence of imaging losses ($\eta=1$, black). However, when one has high imaging losses ($\eta=0.3$, blue), the performance of PTAI is only marginally affected and is better than PCI. $A = 4 \times 4 \mu$m$^2$, $M_\mathrm{p}=75$.}  
  \label{OD3}
\end{figure}

For intermediate ODs ($1 < \beta < 20$), typical of many BEC experiments, the value of $\eta$ becomes important when comparing the two techniques. In this range, PTAI is typically quantum projection noise limited and therefore less sensitive to imaging losses, which often arise due to the complexity of ultracold gas experiments where multiple beams are folded along the imaging path with beamsplitters. In such a scenario, as is the case of our experimental apparatus~\cite{CriticalVelocity} where $\eta \approx 0.3$, as shown in Fig.~\ref{OD3}, PTAI performs better for $\beta=2$ (in the absence of imaging losses, \textit{i.e.} $\eta=1$, PCI performs better).

The analysis given and comparisons made so far are based on the fundamental detection limits due to shot noise in the imaging process. In reality, there are other technical noise sources, such as detector noise, laser noise, imperfect image normalization due to optical fringes, \textit{etc.} that also affect the measurement uncertainty. While most noise sources are common to both imaging techniques, the variation in the transfer fraction $f$ due to technical noise in PTAI has no equivalent analog in PCI. In using a microwave pulse on a single-photon transition for the partial transfer, as in our system, the uncertainty in $f$ can arise from fluctuations in the pulse area or the frequency of the pulse (with respect to the transition resonance). In our system, these fluctuations were sufficiently low not to affect our measurement (section \ref{sec:Case2}). For the cases shown in Fig.~\ref{ODLowHigh}(b) and Fig.~\ref{OD3}, a $<8$\% uncertainty in $f$, which requires system stability that is typically achievable for experiments, is low enough for PTAI to be advantageous. 

\section{Conclusion}

In conclusion, we have discussed and evaluated the technique of partial-transfer absorption imaging.  PTAI extends absorption imaging into the minimally-destructive regime in a way that enables one to perform measurements over the entire range of optical depth. 

 We have discussed the scope and applicability of the technique and demonstrated some of its practical uses in experiment. We have presented a simple model for calculating the measurement uncertainty, which we use to illustrate the advantages of PTAI over standard absorption imaging and compared PTAI to phase-contrast imaging as a minimally-destructive technique. While at low optical depth, PTAI performs better than PCI, for most cases, it gives comparable or slightly poorer performance, but is less sensitive to imaging losses. Furthermore, PTAI has the advantage of easily altering the transferred OD from image to image, allowing optimal imaging of a large dynamic range of OD with no change to the optical setup.

PTAI is a versatile technique with several applications. PTAI can be used to monitor an atomic sample and compensate for unwanted variations in experimental conditions such as atom number or density. An example of a non-imaging application of PTAI is to suddenly remove a deterministic fraction of atoms from the sample (homogeneously), which could take the sample to a non-equilibrium state. This could cause shape oscillations\cite{KetterleCollectiveExcitations} or could induce a phase-transition. When used with Raman transfer, PTAI could be coupled with some momentum or phase analysis, as was done in measuring the coherence length of a quasi-2D cloud~\cite{BKTNIST}.

\section{Acknowledgements}

We would like to thank S. L. Rolston and W. T. Hill for helpful discussions, and R. B. Blakestad for a careful reading of the manuscript. This work was partially supported by ONR, the ARO atomtronics MURI, and the NSF PFC at JQI. 


\appendix 

\section{Calculation of PTAI measurement uncertainty} 
\label{sec:PartialTransferCalc}


In this section, we calculate the measurement uncertainty in PTAI, which is equation (\ref{eq:PTdBr}) in the text. Starting from the expression for the total noise in the number of detected photons (equation (\ref{eq:totalnoise})), we calculate the fractional uncertainty $\delta \beta/ \beta$ in the (on-resonant) OD $\beta$ in the area of interest $A$, which is determined from the measurement of $\eta \bar{M}_t$. In using PTAI, we choose a transfer fraction $f$, giving an average transferred OD of $f \beta$. From the measured photon counts on the detector $\eta \bar{M}_t$, we determine the transferred OD $f \beta$, and hence $\beta$:
\begin{equation}
   \beta = - \frac{1}{f}\log{ \frac{\eta \bar{M}_t}{\eta \bar{M}}}. 
\end{equation}
We assume that the reference signal, $\eta \bar{M}$, which comes from the image of the probe beam on the CCD in the absence of atoms, has been averaged over several realizations and has no uncertainty associated with it. Correspondingly, the uncertainty $\delta \beta$ can be expressed in terms of the uncertainty of the transmitted probe light. 
\begin{equation}
\delta \beta = \frac{1}{f} \frac{\delta \mathcal{N}}{\eta \bar{M_t}}, \label{eq:Uncertainty}
\end{equation}
where $\delta \mathcal{N}$ is the noise on the detected photons $\eta \bar{M}_t$. Using equation (\ref{eq:transmission}) and the optical-pumping-limited number of incident photons (given in equation (\ref{eq:photonlimit})), we find the transmitted number of photons:
\begin{equation}
   \bar{M_t} = \frac{M_\mathrm{p} f N e^{-f \beta}}{1- e^{-f \beta}}, \label{eq:UncertaintyDenom}
\end{equation}
where $M_\mathrm{p}$ is the number of photons an atom can absorb before optical pumping affects the measurement.
The variation in the detected photon counts is given by eq. (\ref{eq:totalnoise}):
\begin{equation}
   \delta \mathcal{N} = \sqrt{\frac{M_\mathrm{p} f N \eta e^{-f \beta}}{1- e^{-f \beta}} \left( 1 + \frac{M_\mathrm{p} \eta (f \beta)^2 e^{-f \beta}}{1-e^{-f \beta}} \right)}. \label{eq:UncertaintyNum}
\end{equation}
Substituting equations (\ref{eq:UncertaintyDenom}) and (\ref{eq:UncertaintyNum}) into equation (\ref{eq:Uncertainty}), and dividing by $\beta$, we find an expression for the fractional uncertainty:
\begin{equation}
   \frac{\delta \beta}{\beta} = \frac{1}{f \beta} \sqrt{\frac{1-e^{-f \beta} +M_\mathrm{p} \eta (f \beta)^2 e^{-f \beta} } {M_\mathrm{p} \eta f \beta (A/\sigma_0) e^{-f \beta} } }  \label{eq:PTIUncFinal},
\end{equation}
where we have substituted $N = \beta A/\sigma_0$, where $\sigma_0$ is the absorption cross-section. This is given as equation (\ref{eq:PTdBr}) in the text. Equation (\ref{eq:PTIUncFinal}) expresses the fractional uncertainty in the measured value of $\beta$ as a function of the OD ($f \beta$) of the transferred fraction, for a given area of interest $A$ and detection efficiency $\eta$. Figures \ref{ODLowHigh} and \ref{OD3} illustrate these dependences. 

\section{Phase-Contrast Imaging measurement uncertainty} \label{sec:PhaseContrastCalculations}

In this section, we derive the expression for the fractional uncertainty in a phase-contrast imaging measurement, which is equation (\ref{eq:PCdBrSimpl}) in the text. We choose a region of interest identical to the PTAI case, having $N$ atoms within area $A$ as shown in Fig.~\ref{ImagingCell} and an (on-resonant) OD $\beta$. In PCI, the probe light is far detuned from resonance, $\Delta \gg 1$. Applying this limit to equations (\ref{eq:transmission}, \ref{eq:absorption} and \ref{eq:phaseshift}), we obtain:
\begin{eqnarray}
\bar{M}_t &=& \bar{M} e^{-\beta / \Delta^2} \\
\bar{M}_\mathrm{abs} &=& \bar{M} (1- e^{-\beta / \Delta^2}) \label{eq:PCI_abs} \\
\delta \phi &=& \frac{\beta}{2 \Delta}. \label{eq:phaseshiftsimpl}
\end{eqnarray}

To retain sensitivity at high ODs, one typically sets $\delta \phi < \pi /4 $, which requires $\Delta > 2 \beta/\pi$. This condition ensures that $e^{-\beta / \Delta^2} \approx 1 - \beta/ \Delta^2$, and the absorption of photons through the sample can be well approximated by:
\begin{equation}
	\bar{M}_\mathrm{abs} = \bar{M} \frac{\beta}{\Delta^2} . \label{eq:absSimpl}
\end{equation}
Since $\bar{M}_\mathrm{abs} \ll \bar{M}$, we can neglect the attenuation of the probe beam, i.e. $\bar{M}_t \approx \bar{M}$.

\subsection{Phase measurement using a local oscillator}

To measure the phase-shift of the transmitted probe beam, we assume a model interferometer in the spirit of  Lye \textit{et al.}\cite{NDI1}, where the transmitted probe beam, $\bar{M}$ interferes with a reference local oscillator that delivers $\bar{M}_L$ photons to the detector during the probe time $\tau$~\cite{twoportinterferometry}. This model is also a reasonable description of more typical PCI setups~\cite{KetterlePhaseContrast}, where one uses a single, large-diameter probe beam. There, most of the probe does not pass through the sample and serves as the local oscillator.

To account for the detection efficiency of both probe and local oscillator in our formalism, we express the efficiency coefficient $\eta$ as a product of the transmission efficiency $\eta_t$ of the probe optical path and the quantum efficiency $\eta_q$ of the detector:
\begin{equation}
 \eta = \eta_t \eta_q.
\end{equation}
For our model interference setup, the detected number of photons is:
\begin{equation}
  \bar{M}_{d} = \eta_q [\bar{M}_L + \eta_t \bar{M} + 2 \sqrt{\bar{M}_L \eta_t \bar{M}} \cos(\phi_0 + \delta \phi),] \label{eq:Interference}
\end{equation}
where $\phi_0$ is the phase shift between the local oscillator and the probe beam in the absence of atoms. For maximum sensitivity, we set $\phi_0 = \pi/2$~~\cite{phasespot}. Equation (\ref{eq:Interference}) becomes:
\begin{equation}
 \bar{M}_{d} = \eta_q [\bar{M}_L + \eta_t \bar{M} - 2 \sqrt{\bar{M}_L \eta_t \bar{M}} \sin(\delta \phi) ]. \label{eq:PCIdet}
\end{equation}

In the absence of atoms, $\delta \phi=0$ and the number of photons incident on the detector is:
\begin{equation}
 \bar{M}_{d_0} = \eta_q (\bar{M}_L + \eta_t \bar{M}). \label{eq:PCIdet_noatoms}
\end{equation}

Substituting the phase shift $\delta \phi$ from equation (\ref{eq:phaseshiftsimpl}) into equation (\ref{eq:PCIdet}), we obtain the detected number of photons, $\bar{M}_{d}$, in the presence of atoms:
\begin{equation}
   \bar{M}_{d} = \eta_q \left[ \bar{M}_L + \eta_t \bar{M} - 2 \sqrt{\bar{M}_L \eta_t \bar{M}} \sin{\left(\frac{\beta}{2 \Delta}\right)} \right]. \label{eq:B9}
\end{equation}
The photon shot noise is the square root of the number of photons incident on the detector, $\bar{M}_{d}$:
\begin{equation}
\delta \mathcal{N}_\mathrm{ph} = \sqrt{\eta_q \left[ (\bar{M}_L + \eta_t \bar{M}) - 2 \sqrt{\bar{M}_L \eta_t \bar{M}} \sin{\left(\frac{\beta}{2 \Delta}\right)} \right] } \label{eq:PCshotnoise}
\end{equation}

\subsection{Uncertainty in the measured Optical Depth}

  As in the PTAI case, we start with an expression for the measured OD $\beta$, and obtain an expression for the uncertainty $\delta \beta$ arising from the photon shot noise, $\delta \mathcal{N}_\mathrm{ph}$.  Subtracting equation (\ref{eq:PCIdet_noatoms}) from equation (\ref{eq:B9}) we obtain an expression for the measured OD:
\begin{equation}
    \beta = 2 \Delta \arcsin \left( \frac{\bar{M}_{d_0} - \bar{M}_{d}}{2 \eta_q \sqrt{\bar{M}_L \eta_t \bar{M}}} \right),
\label{eq:PCOD}
\end{equation}
    where $\eta_q \bar{M}_L$, $\eta_q \eta_t \bar{M}$ and $M_{d_0}$ are measured prior to taking the image.  We assume that these three quantities have been measured repeatedly and so have negligible shot noise associated with them.  

Differentiating equation (\ref{eq:PCOD}) with respect to the signal, $\bar{M}_d$, we get:
\begin{equation}
   \delta \beta = \left[ 1 - \left( \frac{\bar{M}_{d_0} - \bar{M}_{d}}{2 \eta_q \sqrt{\bar{M}_L \eta_t \bar{M}}} \right)^2 \right]^{-\frac{1}{2}} \! \! \! \! \frac{ (\delta M_{d}) \ \Delta}{\eta_q \sqrt{\bar{M}_L \eta_t \bar{M}}}
\end{equation}
   Substituting the variation in photon number, $\delta M_d = \delta \mathcal{N}_\mathrm{ph}$ from equation (\ref{eq:PCshotnoise}) and simplifying the term in the square brackets using equation (\ref{eq:PCOD}) and the identity $(1-\sin^2 x) = \cos^2 x $, the measurement uncertainty is:
\begin{equation}
   \delta \beta = \Delta \frac{\sqrt{\eta_q \left[(\bar{M}_L + \eta_t \bar{M}) - 2 \sqrt{\bar{M}_L \eta_t \bar{M}} \sin{\left(\frac{\beta}{2 \Delta}\right)} \right]} }{{\cos{\left(\frac{\beta}{2 \Delta}\right)}} \eta_q \sqrt{\bar{M}_L \eta_t \bar{M}}}.  
\label{eq:PCI_dBeta}
\end{equation}

   Taking the local oscillator to be much greater than the signal ($\bar{M}_L >> \eta_t \bar{M}$), equation (\ref{eq:PCI_dBeta}) simplifies to
\begin{equation}
   \delta \beta = \Delta \ \frac{1}{\cos{\left(\frac{\beta}{2 \Delta}\right)}} \sqrt{\frac{1}{\eta \bar{M}}},
\end{equation}
where $ \eta = \eta_t \eta_q$.

This assumes a sufficient dynamic range of the detector. For a given $\beta$, the measurement uncertainty depends only on the probe detuning, the number of incident photons.
 
  Although we only measure the phase-shift of the probe beam due to the sample, there is some absorption of probe light. Expressing $\bar{M}$ in terms of $\bar{M}_\mathrm{abs}$ using equation (\ref{eq:absSimpl}), we find:
\begin{equation}
   \delta \beta = \frac{1}{\cos{\left(\frac{\beta}{2 \Delta}\right)}} \sqrt{\frac{\beta}{\eta \bar{M}_\mathrm{abs}}}.
\end{equation}

   Each absorbed photon corresponds to an atom undergoing a recoil event. We can set $\bar{M}_\mathrm{abs}=f_\mathrm{r} N$, where $f_\mathrm{r}$ is the fraction of atoms undergoing recoil events, giving 
\begin{equation}
   \delta \beta = \frac{1}{\cos{\left(\frac{\beta}{2 \Delta}\right)}} \sqrt{\frac{\beta}{\eta f_\mathrm{r} N}}, 
\end{equation}
where the uncertainty of the measurement has been expressed in terms of the perturbation to the sample. This is given as equation (\ref{eq:PCdBrSimpl}) in the text.


\begin{thebibliography}{37}
\bibitem [{sta()}]{starkshift}%
  \BibitemOpen
  \href@noop {} {}\bibinfo {note} {While there is a spatially dependent AC
  Stark shift in optical dipole traps, the differential stark shift between $|g
  \rangle$ and $|a \rangle$ is typically small.}\BibitemShut {Stop}%
\bibitem [{rep()}]{repumper}%
  \BibitemOpen
  \href@noop {} {}\bibinfo {note} {There is a possibility of using a second
  optical repumping beam to bring the optically pumped atoms back to the
  cycling transition. However, in alkali atoms, the transferred fraction is
  usually optically pumped back to $|g \rangle$, and any repumping beam will
  affect the untransferred atoms also.}\BibitemShut {Stop}%
\bibitem [{Cyc()}]{CyclingTransition}%
  \BibitemOpen
  \href@noop {} {}\bibinfo {note} {If one were to use a suitable circular
  polarization of light to excite specifically to the m$_F$ = $\pm$3 magnetic
  sublevels of the F$'$=3 state, spin selection rules further inhibit
  off-resonant excitation to F$'$=2.}\BibitemShut {Stop}%
\bibitem [{N_f()}]{N_fvsN_d}%
  \BibitemOpen
  \href@noop {} {}\bibinfo {note} {$f$ and $f_\mathrm{r}$ are typically
  equivalent for quatum degenerate clouds. Both correspond to the fraction of
  atoms that leave the cloud due to the high kinetic energy (compared to other
  energy scales) acquired by the absorption and reemission of photons during
  the imaging process. However, it should be noted that atoms leaving the cloud
  after a PTAI image have many single photon recoil momenta, while the
  corresponding atoms in PCI or minimally-destructive absorption imaging only
  have $\sqrt 2$ recoil momenta on average.}\BibitemShut {Stop}%
\bibitem [{two()}]{twoportinterferometry}%
  \BibitemOpen
  \href@noop {} {}\bibinfo {note} {To conserve energy, an optical
  interferometer must necessarily have two output ports. Here, we assume that
  the combining optic is designed such that nearly all the transmitted probe
  light is directed to the output port where our detector is
  placed.}\BibitemShut {Stop}%
\bibitem [{pha()}]{phasespot}%
  \BibitemOpen
  \href@noop {} {}\bibinfo {note} {Shifting the local oscillator by $\pi/2$ is
  equivalent to using a phase-spot to shift the phase of the unperturbed probe
  by $\pi/2$ as was done in Ref. \cite{KetterlePhaseContrast}.}\BibitemShut
  {Stop}%
\end{thebibliography}
\end{document}